# Adiabatic frequency shifting in epsilon near zero materials:
# The role of group velocity

Jacob B. Khurgin,[1*] Matteo Clerici,[2] Vincenzo Bruno,[3] Lucia Caspani,[4] Clayton DeVault,[5] Jongbum Kim,[6] Amr Shaltout,[6,7] Alexandra Boltasseva,[6] Vladimir M. Shalaev,[6] Marcello Ferrera,[8] Daniele Faccio,[3] Nathaniel Kinsey,[9]

[1] Dept. of Electrical and Computer Engineering, Johns Hopkins University, Baltimore 21218, USA
[2] School of Engineering, University of Glasgow, G12 8LT Glasgow, United Kingdom
[3] School of Physics and Astronomy, University of Glasgow, Glasgow G12 8QQ, United Kingdom
[4] Institute of Photonics, Department of Physics, University of Strathclyde, Glasgow, United Kingdom
[5] Dept. of Physics and Astronomy, Purdue University, West Lafayette, IN 47907, USA
[6] School of Electrical and Computer Engineering, West Lafayette, IN 47907, USA
[7] Now at Geballe Lab for Advanced Materials, Stanford University, Stanford, CA 94305
[8] Institute of Photonics and Quantum Sciences, Heriot-Watt University, SUPA, Edinburgh EH14 4AS, United Kingdom
[9] Dept. of Electrical and Computer Engineering, Virginia Commonwealth University, Richmond, VA 23284, USA

**The conversion of a photon's frequency has long been a key application area of nonlinear optics. In addition to traditional second- and third-order wave-mixing effects, the adiabatic frequency shift (AFS) has recently been investigated by inducing a time-varying refractive index change in the host material or structure. Here, we investigate the AFS process in epsilon near zero (ENZ) materials and show that while the maximum frequency shift for a given change of permittivity does not exhibit an increase in the vicinity of Re{$\varepsilon_r$}=0, the given permittivity modulation can be achieved in a shorter length and, if the pump is also in the ENZ region, at a lower pump intensity. This arises due to slow propagation effects and results in a photon conversion percentage that is comparable to micro-resonators and other structured slow light schemes. Moreover, no nanofabrication is required for ENZ materials which constitutes their major advantage over alternative frequency conversion approaches. Our results, supported by experimental measurements, indicate that transparent metal oxides operating near the ENZ point are good candidates for future frequency conversion schemes.**

While the field of frequency conversion is well established, the development of powerful ultra-fast light sources has led to the observation of different effects where the entire spectrum of the incoming light wave shifts adiabatically in the presence of a time-varying refractive index [3], typically induced by a strong optical pump (which does not preclude any other means of changing the permittivity, say electro-optic or acousto-optic). These effects have been observed both in high-Q resonant dielectric structures (Fig. 1a) [4-7] and in free propagating waves inside nonlinear epsilon-near-zero (ENZ) media (Fig. 1b) [8-12]. Although several nonlinear processes have been shown to be enhanced in ENZ materials [13-15], driving intense research [16,17], rapid and complete frequency conversion or modulation represents a unique opportunity to explore novel applications, especially in imaging, sensing, and telecommunications. Currently, the research on adiabatic frequency shift (AFS) is ongoing and it is not clear which approach (i.e. high-Q resonators or ENZ) is the most appropriate. With that in mind, we explore what characteristics are needed to achieve efficient AFS, giving special attention to ENZ materials, where we find the potential for large frequency shifts without nanostructuring, at reduced pump intensities, and at smaller

## INTRODUCTION

Frequency conversion is a signature characteristic and a key application of nonlinear optics [1]. It can be attained using both second and third-order nonlinearities in various media. Second-order processes require non-centrosymmetric media and include the well-known effects of sum and difference frequency generation while third-order phenomena, which occur in any medium, include four-wave mixing processes as well as Raman effects. All of these frequency conversion schemes can be characterized as parametric processes where the optical properties are modulated with a certain frequency which causes the generation of oscillations at new frequencies. Typically, frequency conversion is well described by the coupled-wave formalism [2] in which the energy flows back and forth between waves of different frequencies (often characterized as pump, signal, and idler) depending on the phase relations between the waves (phase-matching conditions).

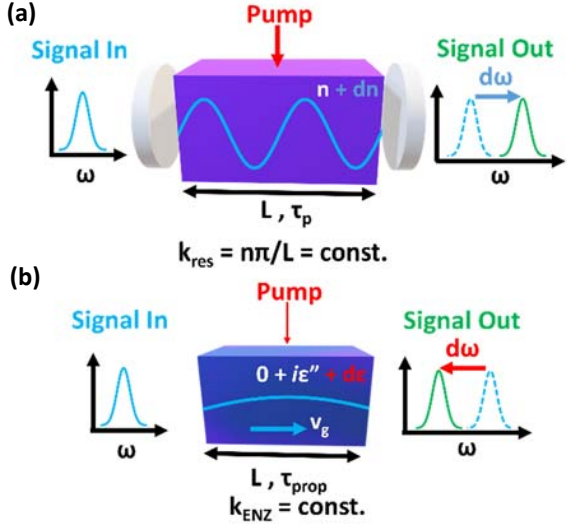

**Fig. 1.** Adiabatic frequency shift in **(a)** microcavity (or a photonic crystal) and **(b)** unconstrained epsilon-near-zero medium in which a rapid change of refractive index δn (and permittivity δε) causes a frequency shift δω.

lengths compared to previous schemes (Fig. 1). In this, we focus on the role of the group velocity in the development of the AFS process, drawing connections between AFS in ENZ and fields of slow and fast light.

## ADIABATIC FREQUENCY SHIFT IN ENZ

The underlying cause of AFS due to temporal modulation of the refractive index is rather straightforward: if the refractive index experienced by a probe pulse changes in space or time while crossing the medium, the probe wavevector or frequency will change, respectively. A more general scenario is described in the Supplementary materials "S1. Effects of refractive index changes". Considering the general case of a time-varying refractive index discussed here, we investigate the case of a change in refractive index that occurs on a sufficiently short time scale compared to the probe transit time in the medium ($\tau_{prop}$, see Fig. 1b), such that the change in frequency of the probe is the dominant effect. Within this case, an adiabatic frequency shift can occur (see e.g. Ref [3]) depending on the duration of the refractive index change compared to the probe pulse duration. More specifically, the refractive index must change over a time scale $\Delta t_{change}$ that is longer than the inverse of the pulse bandwidth $\Delta\omega_{probe}$ and the inverse of the eigenmode spacing $\Delta\omega_{em}$, such that only a single mode remains excited and no modulation sideband frequencies are generated within the pulse. In our case (free space propagation), this is simply given by $\Delta t_{change} > \tau_{probe}$ where $\tau_{probe}$ is the probe pulse width.

Similar arguments can be made for resonant structures, wherein a sufficiently quick change in the index ($\tau_p \gg \Delta t_{change} \gg \tau_{rt}$, where $\tau_p$ is the photon lifetime in the cavity and $\tau_{rt}$ is the cavity round trip time) gradually modifies the resonant condition ($k_{res} = n\pi/L$) and therefore generating phase and amplitude modulation (see Fig. 1a).

It is important to mention that the microscopic mechanisms that support AFS are the same that underpin parametric processes, e.g. four-wave mixing (FWM), taking place in more conventional frequency conversion schemes. In fact, the rapid change of index caused by the pump with a subsequent change in the frequency of the probe can be represented as ultra-broadband FWM by simply applying a Fourier transform, but describing AFS in the time domain is by far more convenient.

From these adiabatic conditions, we then wish to maximize the frequency shift $\delta\omega$ for a given $\delta n$. To accomplish this, one might think to simply differentiate the momentum conservation, or resonance, condition:

$$\omega n(\omega) = k(\omega)c \qquad (1)$$

to obtain

$$\delta(n\omega) = \omega\delta n + n\delta\omega = 0 \qquad (2)$$

from which it follows that

$$\delta\omega = -\frac{\delta n}{n}\omega. \qquad (3)$$

In other words, if the material has a very small index then a small change would lead to a very large adiabatic frequency shift. With this in mind, we consider the permittivity for ENZ materials that can be written as:

$$\varepsilon_r(\omega) = \varepsilon_\infty\left(1 - \frac{\omega_p^2}{\omega^2 + i\omega\gamma}\right), \qquad (4)$$

where $\varepsilon_\infty$ is the permittivity associated with the valence electrons (band-to-band transitions), $\omega_p = Ne^2/\varepsilon_0\varepsilon_\infty m^*$ is the plasma frequency, $N$ is the free carrier density, $m^*$ is the effective mass, and $\gamma$ is the loss rate (which we shall omit for the sake of simplicity here until comparing with experimental results). As seen from Eq. (4), the permittivity and index of refraction become small in the vicinity of $\omega_p$, giving rise to an ENZ region. By differentiating the index ($n(\omega) = \varepsilon_r(\omega)^{1/2}$), the change in the ENZ material's index can be described by:

$$\delta n(\omega) = \delta\varepsilon(\omega)/2n(\omega) \qquad (5)$$

and is found to be quite large since $n(\omega) \to 0$. By substituting Eq. (5) into Eq. (3), the resulting frequency shift can be described as:

$$\delta\omega = -\frac{\delta\varepsilon_r}{2n^2}\omega = -\frac{\delta\varepsilon_r(\omega)}{2\varepsilon_r(\omega)}\omega. \qquad (6)$$

Equation (6) states that for a given change of the material's permittivity, the corresponding frequency shift is enhanced in an ENZ environment. However, the robustness of this result will be tested in the following section.

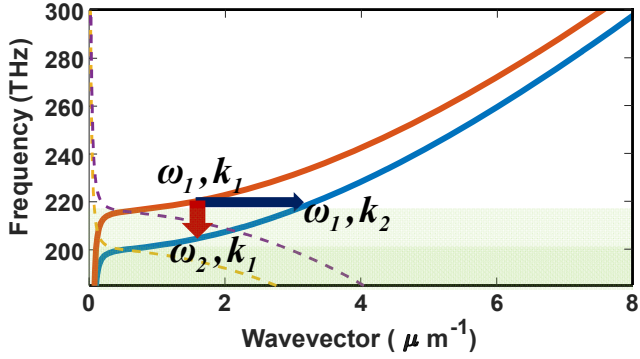

**Fig. 2**. Adiabatic frequency shift in the material with Drude-like dispersion. As the permittivity changes, the dispersion curve shifts from 1 to 2. If the change of permittivity is very slow, the frequency $\omega_1$ stays unchanged while the wave vector shifts rightward from $\mathbf{k}_1$ to $\mathbf{k}_2$. If the change in permittivity is sufficiently fast, then the wavevector $\mathbf{k}_1$ remains unchanged while the frequency shifts downward from $\omega_1$ to $\omega_2$. Also shown are imaginary parts of wavevector (dashed lines) and the tinted regions show the high reflectivity (or "metallic") region with a negative real part of permittivity.

## GROUP VS. PHASE VELOCITY

From Eq.(6), it appears then that an ENZ material is an ideal medium for AFS as has indeed been confirmed by the experimental results [10]. However, a quantitative analysis of AFS in ENZ systems based on Eq. (2), might lead to considerable errors because the adiabatic shift $\delta\omega$ is obtained using only a partial and not a full derivative. A more accurate way to evaluate the refractive index change should account for both the dispersion and the index change experience by the probe wave as it propagates in the pumped medium as:

$$n'(\omega + \delta\omega) - n(\omega) \approx \delta n(\omega) + \delta\omega \frac{dn(\omega)}{d\omega}. \quad (7)$$

Here $n'$ indicates the refractive index after the modulation has occurred, $\delta n(\omega)$ is the change on the linear refractive index imposed by the external pump (typically optical pulse), and the last term represents the strong dispersion of the material near ENZ. If Eq. (7) is combined with the condition of momentum conservation in Eq. (1) we find:

$$\omega n(\omega) \approx (\omega + \delta\omega)\left[ n(\omega) + \delta n(\omega) + \delta\omega \frac{dn(\omega)}{d\omega} \right]. \quad (8)$$

From Eq. (8), where the full derivative is taken into consideration, the induced frequency shift is:

$$\delta\omega = -\frac{\delta n(\omega)}{n_g(\omega)}\omega, \quad (9)$$

where $n_g$ is the group index given by:

$$n_g(\omega) = n(\omega) + \omega \frac{dn}{d\omega}. \quad (10)$$

In this sense, we may also interpret the adiabatic frequency shift as the rate at which the additional phase accumulation over the propagation distance $z(t) = v_g t$ changes due to the nonlinear index modulation, (where $v_g = c/n_g$ is the group velocity):

$$\delta\omega(t) = -\frac{\partial}{\partial t}\delta\Phi(t) = -\frac{\omega}{c}\frac{\partial}{\partial t}\int_0^{z(t)} \delta n(z)dz =$$
$$= -\frac{\omega}{c}\frac{\partial}{\partial t}\int_0^t v_g \delta n(t')dt' = -\frac{\omega}{n_g}\delta n(t) \quad (11)$$

Using this description, we see that Eq. (11) and Eq. (9) are identical, and that the maximum frequency shift is achieved at a time $t_{max}$ after the pulse enters the material given by $\Delta t_{change}$ or the propagation time $\tau_{prop} = L/v_g$, whichever is shorter. Of course, the presence of group rather than phase velocity in the expression for AFS also allows from the interpretation of AFS as a Doppler shift [21].

Now, the group index in ENZ materials is actually very large and can be estimated by multiplying both sides of Eq. (10) by $n(\omega)$ and then using Eq. (4):

$$n(\omega)n_g(\omega) = \varepsilon_r(\omega) + \frac{1}{2}\omega\frac{d\varepsilon_r(\omega)}{d\omega} = \varepsilon_\infty. \quad (12)$$

Substituting Eq. (5) and Eq. (12) into Eq. (9) gives:

$$\delta\omega \approx -\frac{\delta\varepsilon_r(\omega)}{2\varepsilon_\infty}\omega \quad (13)$$

which suggests that no ENZ enhancement for AFS is expected, in opposition to the findings of Eq. (6). However, this conclusion cannot explain the large frequency shift achieved in [8-10]. So, let us dig deeper.

The change in the permittivity for ENZ materials can be induced by varying the carrier density, the average effective mass, and to a lesser extent the "background" dielectric constant $\varepsilon_\infty$. In other words, it amounts to modifying the plasma frequency of the material, i.e. the frequency at which the ENZ condition is achieved. When the frequency of the pump exceeds the bandgap of the ENZ material, such as ITO or AZO, carriers are excited into the conduction band via interband absorption, thereby causing a blue-shift of the plasma frequency, which relaxes through recombination processes [13,14]. However, if the energy of the pump is below the bandgap, the interaction tends to red-shift the plasma frequency. This occurs because the intraband absorption process elevates free-electrons in the conduction band to high-energy states which have a larger effective mass than states near the bottom of the band due to a non-parabolic dispersion [18]. These carriers then relax through scattering processes lasting on the order of a few 100's of femtoseconds [19], much faster than interband relaxation which can extend to one nanosecond (although this time can also be significantly shortened [2,13]). For more information on the microscopic processes occurring in interband and intraband nonlinearities of ENZ materials, we direct the reader to [16-19].

From the realization that the nonlinearity can be reduced to a modification of the plasma frequency, Eq. (4) may be differentiated against the induced plasma frequency change, while neglecting the damping factor, which yields:

$$\delta\varepsilon_r(\omega) = -2\varepsilon_\infty \omega_p \delta\omega_p / \omega^2. \quad (14)$$

Substituting this into Eq. (13) yields a strikingly simple expression:

$$\delta\omega \approx \delta\omega_p \frac{\omega_p}{\omega} \approx \delta\omega_p. \quad (15)$$

In other words, the frequency of the probe beam simply shifts with the entire dispersion curve, described by:

$$k(\omega) = \frac{\omega}{c} n(\omega) = \frac{\omega}{c}\sqrt{1-\frac{\omega_p^2}{\omega^2}} = c^{-1}\varepsilon_\infty^{1/2}\sqrt{\omega^2-\omega_p^2}. \quad (16)$$

In fact, by enforcing a constant wavevector to Eq. (16), Eq. (15) is obtained, indicating that the frequency of the probe beam follows the plasma frequency as shown in Fig. 2. For intraband nonlinearities, the reduction of the plasma frequency causes a downward shift of the dispersion curve, producing an adiabatic frequency shift (downward arrow) that is relatively constant in frequency as predicted by Eq. (15). This is strikingly different from the phase shift enhancement in the ENZ material. In the case of a very slow index change where the frequency is preserved but the wavevector changes as described in Fig. S1(a), a large momentum (and phase) change is observed in the ENZ region for a small change of the dielectric constant (see Fig. 2 horizontal arrow). This occurs due to the minimal slope of the dispersion curve as the frequency approaches ENZ.

## REAL BENEFIT OF ENZ MATERIALS

So, how can one explain the impressive adiabatic frequency shifts obtained using ENZ materials [8-10]? The answer brings us back to the propagation expression in Eq. (11). From here, it follows that while the value of the frequency shift $\delta\omega$ for a given $\delta\varepsilon$ does not change as a function of the group velocity, the length over which the shift occurs ($L_{change} \sim v_g \Delta t_{change}$) is shorter near the ENZ condition due to the slow propagation velocity. Therefore, even if the actual sample length $L$ is shorter than $L_{change}$, as is the case in the experiments detailed below, a large index change $\delta n$ will be seen by the probe. This represents a significant advantage, giving rise to three benefits: 1) the potential to use thinner materials, 2) an improved temporal overlap of the light with slower and therefore stronger nonlinearities (as was employed in [8,9]), and 3) a reduction in the pump intensity for a given $\delta\varepsilon$, because $\delta\varepsilon \sim \chi^{(3)} E_{pump}^2$ and the pump energy is compressed in the region of slow propagation $E_{pump}^2 \sim 2\eta_o n_g I_{pump}/\varepsilon_\infty$, where $I_{pump}$ and $\eta_o$ are the pump intensity and the vacuum impedance, respectively. It is important to note that this enhancement is not intrinsic to ENZ materials as it is entirely due to slow propagation. In fact, the enhancement of third-order nonlinear effects by $n_g^2$ occurs in all structural (photonic) slow light schemes [22-24] including micro-resonators [7], where one can simply use cavity finesse in place of $n_g$ to estimate the enhancement. The inherent benefit of ENZ materials is that they offer a similar enhancement over a shorter length that requires no intricate nanofabrication whatsoever, which in our eyes is an enormous advantage in practical applications [19]. Of course, these advantages need to be weighed against the fact that the nonlinearity in ENZ materials is entirely due to absorption, i.e. the scheme is lossy, but as experimental results demonstrate, a high conversion percentage can be achieved (defined as the power of AFS beam divided by the incident probe power).

From this understanding, let us define the figure of merit for AFS as:

$$FOM = -\frac{\delta\omega}{\omega I_{pump} L} = \frac{\chi^{(3)\prime}\Delta t_{change} 2\eta_0 n_g I_{pump}/\varepsilon_\infty}{2\varepsilon_\infty I_{pump} L} =$$
$$= n_g^2 \frac{\chi^{(3)\prime}\mu_0}{\varepsilon_\infty^2} \approx \frac{\chi^{(3)\prime}\mu_0}{n^2} = \chi^{(3)\prime}\varepsilon_0\eta^2 \quad (17)$$

Here, we use the fact that the third-order susceptibility, when it originates from the excitation of real carriers like in ITO or AZO, can be written as a product of interaction time $\Delta t_{change}$ and the derivative of susceptibility with respect to time, $\chi^{(3)\prime}$. The latter depends only on the matrix elements of the optical dipole for intraband transitions, and also (as is the case for ENZ) on the nonparabolicity of the conduction band. This derivative does not change significantly from one nonlinear material to another which explains the well-known fact that slow nonlinearities are stronger than fast processes [4].

Before concluding this theoretical discourse, it is prudent to discuss the region of very small wave-vectors in Fig. 2 where the dispersion curve "bends down". This is the so-called "fast light region" [25] where $d\omega/dk$ is very large and a small change of permittivity indeed yields a frequency shift larger than $\delta\omega_p$. This is an interesting observation, but it is of not much practical importance for two reasons. First, fast light is always associated with high losses as can be seen by looking at the imaginary wavevector (Fig. 2 dashed lines). Of course, losses may be compensated with gain, but that greatly increases complexity. Second, fast light indicates that no "compression" of energy occurs (in fact just the opposite). Therefore, more pump intensity and/or a "fast light" or "white light" cavity [26] would be needed, which would defeat the purpose of using a simple structure.

It is also worth mentioning that the large impedance mismatch typically associated with ENZ materials naturally limits the input and output energy coupling. Although this has been mitigated in ENZ materials by simultaneously reducing the permeability $\mu_0$ to maintain an impedance match (epsilon- and mu-near-zero materials [27]), as can be seen in Eq. (17) such an approach will ultimately cancel the desired AFS enhancement. This can be realized from another angle as well. Following Eq. (12) the relation between the group index and relative impedance can be obtained, $n_g = \varepsilon_\infty \eta_r$ where $\eta_r = \sqrt{\mu_r/\varepsilon_r}$.

In this sense, ENZ materials do have a weakness of being difficult to couple into (which is no different from photonic slow light structures [5]). However, the magnitude of the mismatch is naturally limited by the losses inherent in the material and can be further mitigated through the excitation of the material at the Brewster angle with p-polarized light, coupling into Brewster or ENZ modes within the material [28], or utilizing a short graded permittivity matching layer.

## EXPERIMENTAL VERIFICATION

The description of AFS in ENZ materials presented has also been experimentally verified with a pump-probe experiment as depicted in Fig. 3a. We find good agreement with the model apart from a correction that will be detailed below.

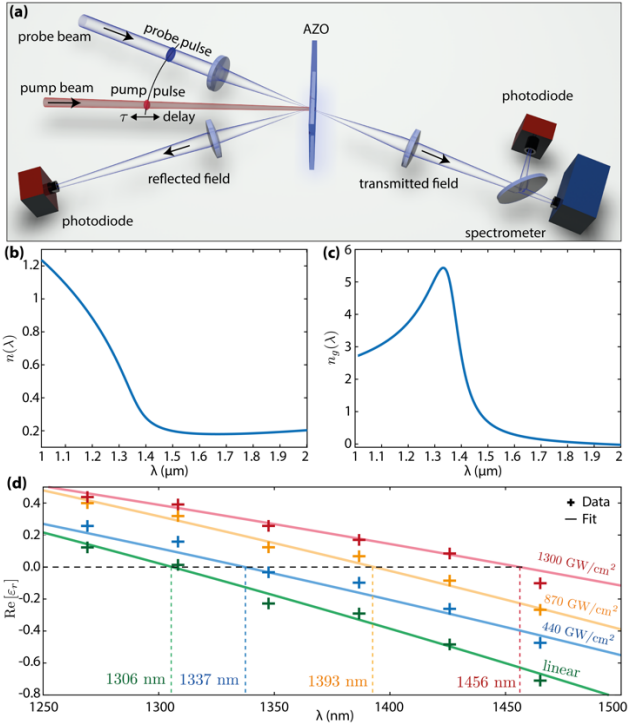

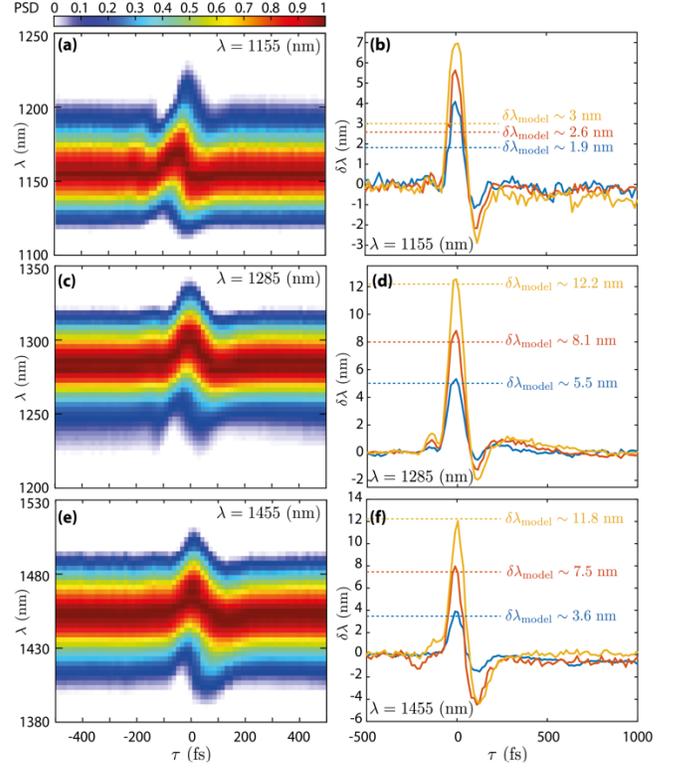

**Fig. 3. (a)** Shows a sketch of the experimental setup. A 785 nm laser pulse train of ~115 fs duration and intensity up to 1.3 TW/cm² excites a 900nm thick AZO film at 100 Hz repetition rate. The reflection and transmission signals are recorded with calibrated photodiodes for a probe pulse with a wavelength that is tuned between ~1250nm and ~1500nm. The spectrum of the transmitted pulse is recorded with a spectrometer. **(b)** Linear index and **(c)** group index of the AZO sample as a function of wavelength. **(d)** Shows the measured values for the real part of the permittivity (crosses) at different pump intensities (green: no pump, blue: ~440 GW/cm², orange: ~870 GW/cm² and red: ~1304 GW/cm²). The solid lined of corresponding colors are obtained from a fit with a Drude model.

The permittivity of a 900 nm thick Oxygen-deprived Aluminum doped Zinc Oxide film (AZO) was characterized from 1150 nm to 1550 nm, in the vicinity of the ENZ region, by recording the reflected ($R$) and transmitted ($T$) power of a probe pulse with calibrated photodiodes. The probe wavelength was tuned by an optical parametric amplifier while maintaining a nearly constant pulse duration of ~100 fs (full width at half-maximum). From the values of $R(\lambda)$ and $T(\lambda)$, the linear values of the $\text{Re}\{\varepsilon_r(\lambda)\}$, shown in Fig. 3b with green crosses, and $\text{Im}\{\varepsilon_r(\lambda)\}$ are extracted using an inverse transfer matrix method [16,29]. The complex permittivity values (the green solid line in Fig. 3b) are then fitted by the Drude model in Eq. (4) to extract $\varepsilon_\infty$, $\omega_p$ and $\gamma$. To improve the fitting conditions, $\omega_p$ and $\gamma$ were first obtained by fitting the ratio $\text{Re}\{\varepsilon_r\}/\text{Im}\{\varepsilon_r\}$. The same procedure is then applied to establish the permittivity and the parameters of the best-fit Drude model in the case of an optically excited sample. In this case, the measured $R$ and $T$ have been performed at the pump-probe delay for which the pump-induced shift of the probe wavelength is maximized.

**Fig. 4. (a) (c)** and **(e)** Show the spectrogram of a probe pulse as a function of the pump-probe delay $\tau$ for three different input central wavelengths (1155 nm, 1285 nm, 1455 nm, respectively) for a pump pulse of $I_p$ ~870 GW/cm² peak intensity. **(b) (d)** and **(f)** show the probe carrier wavelength in function of the pump-probe delay for the three input wavelengths mentioned above, and for three different pump peak intensities (blue: 440 GW/cm², orange: 650 GW/cm², yellow: 870 GW/cm²). The dashed lines are the wavelength shifts predicted by the model, based on the measured $\delta n$ shown in Fig. S2 (Supplementary Materials).

The nonlinear excitation was produced using pump pulses at 785 nm with ~115 fs duration and intensity up to 1.3 TW/cm². We note that in this case, the pump pulse is not in the ENZ region. Three different pump intensities were tested in this experiment, $I_1$ = 440 GW/cm², $I_2$ = 870 GW/cm², and $I_3$ = 1300 GW/cm². The colored crosses show the real part of the permittivity for these three cases in Fig. 3b, along with the fitted Drude dispersion (solid lines). For a lossless medium, the plasma frequency corresponds to the case when $\text{Re}\{\varepsilon_r(\omega_p)\} = 0$. These occur for the linear (un-pumped) case at a wavelength of 1306 nm and for the three optically excited cases at $\lambda_1$ = 1337 nm, $\lambda_2$ = 1393 nm, and $\lambda_3$ = 1456 nm, as shown in Fig. 3d. From this data, it can be clearly seen that the plasma frequency (wavelength in the figure), shifts under the action of an intense pulsed excitation, enabling the potential for AFS as described by Eq. (15).

To predict the frequency shift of the probe, we considered Eq. (9), which estimates the frequency shift based on the measured value of $\delta n$ as opposed to the fitted Drude permittivity. Moreover, this approach holds in the general case where the dispersion is not solely modeled by a Drude relation. We measured time-resolved changes in the probe reflection and transmission as a function of the pump-probe delay $\tau$ for three different probe wavelengths (1155 nm, 1285 nm, and 1455 nm) and at three different pump peak intensities (440 GW/cm², 650 GW/cm², and 870 GW/cm²). Negative $\tau$ represents the case where the probe precedes the pump. From the time-resolved $\Delta R(\tau)$ and $\Delta T(\tau)$ data, shown in Fig. S2 (a)(b)(d)(e)(g)(h) in the Supplementary Materials (S2), we calculated the change in the refractive index $\delta n$ shown in Fig. S2 (c)(f) and (i). We assume that the refractive index change occurs within a time $\Delta t_{change}$ that is the time it takes for the index to go from $n$ to $n + \delta n$. Considering the measurements in Fig. S2, $\Delta t_{change} \sim 50$ fs.

The probe pulse, however, experiences a time-varying medium for the limited time it crosses the sample, $\tau_{prop} = Ln_g / c$. For the three wavelengths considered, these correspond to: $\tau_{prop}$ (1150 nm) ~ 10 fs, $\tau_{prop}$ (1285 nm) ~ 15 fs, and $\tau_{prop}$ (1455 nm) ~ 3.5 fs. Since $\tau_{prop} < \Delta t_{change}$, Eq. (9) should be modified to account for the effect of the short sample by including a scaling term $\tau_{prop} / \Delta t_{change}$. Including this term into Eq. (9), the probe frequency shift is then:

$$\delta\omega \sim -\omega L \delta n / (c \Delta t_{change}) \quad (18)$$

The probe pulse spectrum (power spectral density) as a function of the pump-probe delay for three different probe wavelengths and pump intensity $I_p$ ~870 GW/cm² are shown in Fig 5(a), (c) and (e). Here, we see a clear shift of the entire pulse spectrum, as expected for the AFS process. In Fig 5 (b), (d), (f) the frequency shift of the spectral center of mass at different powers and probe wavelengths is shown as a function of the pump-probe delay. Overlapped, we also show the shift of the wavelength predicted by the model (dashed lines), based on the measured $\delta n$. The model is able to accurately predict the peak value and intensity dependence of the frequency shift for the two longer wavelengths (1285 and 1450 nm) while capturing the trend in the dielectric regime (1155 nm) although the magnitude is underestimated. This discrepancy at 1155 nm is attributed to a reduced signal-to-noise ratio in the measurement of $R$ and $T$, and therefore $\delta n$ which directly impacts the predicted frequency shift magnitude. The good agreement between the predicted and measured peak values confirms the validity of the proposed model with the inclusion of the reduction term that arises in the case $\tau_{prop} < \Delta t_{change}$.

One final point for the AFS in Fig. 5(b)(d),(f) is the small negative shift of wavelength following the pump pulse. This is expected as probe pulses entering the material after the peak change in the index only experience a blue-shift in the plasma frequency, giving rise to a frequency shift of the opposite sign to the excitation. The magnitude of this effect is ~3× weaker because the relaxation is ~3× slower, such that the ratio $\tau_{prop} / \Delta t_{change}$ is reduced.

## CONCLUSIONS

In conclusion, we have developed a simple model for the adiabatic frequency shift that pinpoints the key role played by the group velocity (and not phase velocity as previously believed). Specifically, a reduced group velocity enables the same AFS shift to be achieved in a thinner sample, with less pump energy, and without intricate nanofabrication. This conclusion has been verified experimentally, demonstrating adiabatic shifts of the entire probe pulse spectrum by up to 18 nm and validating the process model with the introduction of a correction factor $\tau_{prop} / \Delta t_{change}$ to account for the short sample. However, it is important to note two final remarks. First, with this short sample correction, the group index of the probe cancels and does not directly manifest in the fitting of the experimental data, see Eq. (18). In this sense, the role of group velocity is somewhat hidden, and ultimately its effect is to enable the AFS process to occur in thin samples and under longer timescales than would generally be observed. Second, the AFS process can be further enhanced by placing the pump within the ENZ regime. In this case, the slow light effect will enhance the electric field of the pump (due to temporal compression) and maximize the $\delta n$ produced in the sample.

The data to support this manuscript are available at DOI: http://dx.doi.org/10.5525/gla.researchdata.954

**Funding.** National Science Foundation (NSF) Award #1741694. M.C. acknowledges funding from UKRI (EP/S001573/1). NK acknowledges funding from the Air Force Office of Scientific Research (AFOSR) FA9550-1-18-0151. MC acknowledges the support from the United Kingdom Research and Innovation (UKRI, Innovation Fellowship EP/S001573/1). MF acknowledges support from EPSRC (UK, Grant EP/P019994/1). JK acknowledges the support of the DARPA NLM program HR00111820063.

**REFERENCES**

1. R. W. Boyd, Nonlinear optics. 3rd ed. (Academic Press: Amsterdam 2008).
2. Y. R. Shen, The Principles of Nonlinear Optics (Wiley Classics Library ed.; Wiley-Interscience 2003).
3. J. Mendonça, "Nonlinear interactions of wave packets," Journal of Plasma Physics, **22**, 15-26 (1979).
4. M. Notomi, S. Mitsugi, "Wavelength conversion via dynamic refractive index tuning of a cavity," Phys. Rev. A **73**, (2006).
5. S. F. Preble, Q. F. Xu, M. Lipson, "Changing the colour of light in a silicon resonator," Nat. Photon. **1**, 293-296 (2007).
6. M. W. McCutcheon, A. G. Pattantyus-Abraham, G. W. Rieger, J. F. Young, " Emission spectrum of electromagnetic energy stored in a dynamically perturbed optical microcavity," Opt. Express **15**, 11472-11480 (2007).
7. T. Tanabe, M. Notomi, H. Taniyama, E. Kuramochi, "Dynamic Release of Trapped Light from an Ultrahigh-Q Nanocavity via Adiabatic Frequency Tuning," Phys. Rev. Lett. **102**, (2009).
8. M. Clerici, N. Kinsey, C. DeVault, J. Kim, E. G. Carnemolla, L. Caspani, A. Shaltout, D. Faccio, V. M. Shalaev, A. Boltasseva, M. Ferrera, " Controlling hybrid nonlinearities in transparent conducting oxides via two-colour excitation," Nat. Commun. **8**, (2017).
9. E. G. Carnemolla, L. Caspani, C. DeVault, M. Clerici, S. Vezzoli, V. Bruno, V. M. Shalaev, D. Faccio, A. Boltasseva, M. Ferrera, " Degenerate optical nonlinear enhancement in epsilon-near-zero transparent conducting oxides," Opt. Mater. Express **8**, 3392-3400 (2018).
10. E. G. Carnemolla, V. Bruno, L. Caspani, M. Clerici, S. Vezzoli, T. Roger, C. DeVault, J. Kim, A. Shaltout, V. Shalaev, A. Boltasseva, D. Faccio, and M. Ferrera, "Giant nonlinear frequency shift in epsilon-near-zero aluminum zinc oxide thin films," in Conference on Lasers and Electro-Optics, OSA Technical Digest (online) (Optical Society of America, 2018), paper SM4D.7.
11. M. R. Shcherbakov, K. Werner, Z. Fan. N. Talisa, E. Chowdhury, G. Shvets, " Photon acceleration and tunable broadband harmonics generation in nonlinear time-dependent metasurfaces," Nat. Commun. **10**, (2019).

12. Y. Yang, J. Lu, A. Manjavacas, T. S. Luk, H. Liu, K. Kelley, J.-P. Maria, E. L. Runnerstrom, M. B. Sinclair, S. Ghimire, I. Brener, "High-harmonic generation from an epsilon-near-zero material," Nat. Phys. **15**, (2019).
13. N. Kinsey, C. DeVault, J. Kim, M. Ferrera, V. M. Shalaev, A. Boltasseva, "Epsilon-near-zero Al-doped ZnO for ultrafast switching at telecom wavelengths," Optica **2**, 616-622, (2015).
14. L. Caspani, R. P. M. Kaipurath, M. Clerici, M. Ferrera, T. Roger, J. Kim, N. Kinsey, M. Pietrzyk, A. Di Falco, V. M. Shalaev, A. Boltasseva, D. Faccio, "Enhanced Nonlinear Refractive Index in ε-Near-Zero Materials," Phys. Rev. Lett. **116**, (2016).
15. M. Z. Alam, I. De Leon, R. W. Boyd, "Large optical nonlinearity of indium tin oxide in its epsilon-near-zero region," Science **352**, 795-797 (2016).
16. O. Reshef, I. De Leon, M. Z. Alam, R. W. Boyd, "Nonlinear optical effects in epsilon-near-zero media," Nat. Rev. Mater. (2019)
17. I. Liberal, N. Engheta, "Near-zero refractive index photonics," Nat. Photon. **11**, 149-159, (2017).
18. H. Wang, K. Du, C. Jiang, Z. Yang, L. Ren, W. Zhang, S. J. Chua, T. Mei, "Extended Drude Model for Intraband-Transition-Induced Optical Nonlinearity," Phys. Rev. Appl. **11**, 064062 (2019).
19. N. Kinsey and J. Khurgin, "Nonlinear epsilon-near-zero materials explained: opinion," Opt. Mater. Express **9**, 2793-2796 (2019).
20. N. Kinsey, C. DeVault, A. Boltasseva, and V. M. Shalaev, "Near-zero-index materials for photonics," Nat. Rev. Mater. **4**, 742-760 (2019).
21. A. Shaltout, M. Clerici, N. Kinsey, R. P. M. Kaipurath, J. Kim, E. G. Carnemolla, D. Faccio, A. Boltasseva, V. M. Shalaev, M. Ferrera, "Doppler-shift emulation using highly time-refracting TCO layer," in Conference on Lasers and Electro-Optics, OSA Technical Digest (online) (Optical Society of America, 2018), paper FF2D.6, 2016.
22. J. B. Khurgin, "Optical buffers based on slow light in electromagnetically induced transparent media and coupled resonator structures: comparative analysis," J. Opt. Soc. Am. B **22**, 1062-1074 (2005).
23. J. B. Khurgin, "Slow light in various media: a tutorial," Adv. Opt. Photonics **2**, 287-318 (2010).
24. J. B. Khurgin, "Performance of nonlinear photonic crystal devices at high bit rates," Opt. Lett. **30**, 643-645 (2005).
25. M. S. Bigelow, N. N. Lepeshkin, R. W. Boyd, "Superluminal and slow light propagation in a room-temperature solid," Science **301**, 200-202 (2003).
26. H. N. Yum, J. Scheuer, M. Salit, P. R. Hemmer, M. S. Shahriar, "Demonstration of White Light Cavity Effect Using Stimulated Brillouin Scattering in a Fiber Loop ," J. Lightwave Technol. **31**, 3865-3872 (2013).
27. A. M. Mahmoud, N. Engheta, "Wave-matter interactions in epsilon-and-mu-near-zero structures," Nat. Commun. **5**, (2014).
28. M. Z. Alam, S. A. Schulz, J. Upham, I. De Leon, R. W. Boyd, "Large optical nonlinearity of nanoantennas coupled to an epsilon-near-zero material," Nat. Photon. **12**, 79-83 (2018).
29. R. M. Kaipurath, M. Pietrzyk, L. Caspani, T. Roger, M. Clerici, C. Rizza, A. Ciattoni, A. Di Falco, D. Faccio, "Optically induced metal-to-dielectric transition in epsilon-near-zero metamaterials," Sci. Rep. **6**, 27700 (2016).